\documentclass[conference]{IEEEtran}			
\usepackage[utf8]{inputenc}
\usepackage{pifont}
\usepackage{float}
\usepackage[english]{babel}
\usepackage{graphicx}
\usepackage{amsmath}
\usepackage{color}
\usepackage{balance}
\usepackage[T1]{fontenc}
\usepackage[ruled,vlined,linesnumbered]{algorithm2e}
\usepackage{url}
\usepackage{booktabs}

\begin{document}

\title{``Shifting Access Control Left'' using Asset and Goal Models}
\author{\IEEEauthorblockN{Shamal Faily}
\IEEEauthorblockA{\textit{Defence Science \& Technology Laboratory} \\
Portsdown West, UK \\
sfaily@dstl.gov.uk}
}
\maketitle

\begin{abstract}
Access control needs have broad design implications, but access control specifications may be elicited before, during, or after these needs are captured.  Because access control knowledge is distributed, we need to make knowledge asymmetries more transparent, and use expertise already available to stakeholders.  In this paper, we present a tool-supported technique identifying knowledge asymmetries around access control based on asset and goal models.  Using simple and conventional modelling languages that complement different design techniques, we provide boundary objects to make access control transparent, thereby making knowledge about access control concerns more symmetric.  We illustrate this technique using a case study example considering the suitability of a reusable software component in a new military air system.
\end{abstract}

\begin{IEEEkeywords}
knowledge asymmetries, access control, UML, KAOS, IRIS, CAIRIS, PYRAMID, DDS
\end{IEEEkeywords}

\section{Introduction}

The UK MOD's Secure by Design programme \cite{sbd} aims to ``shift security left'', to ensure addressing security at the earliest stage of capability acquisition is more than just a platitude, and practical advice for doing this at the pre-concept and concept stages is more than just aspirational.  To achieve this, there is a need for practical techniques and tools to help stakeholders address security challenges early without security getting in the way of innovation.

Compared to authentication - verifying the identity of a user -- there has been comparatively little work on how to effectively design for authorisation - verifying the resources a user can access.  The need for access control is typically expressed at an abstract level, but this is rarely captured adequately in specifications, instead often being encapsulated within detailed designs.  Consequently, because access control mechanisms are rarely perceived directly by end-users, the root causes of access control inconsistency are not always obvious from their symptoms.  As such, if we can build ``access control by design'' not only will this advance complementary security technology relying on fine-grained access control, e.g. data-centric security \cite{wron24}, it will also overcome an important inhibitor to getting stakeholders to think practically about security at the outset of a programme's life.

Access control needs impact the interaction between people and assets, but access control requirements may precede requirements elicitation, be created in parallel by different teams, or ``bolted on'' sometime afterwards.  If a system's access control requirements fail to adequately account for users' access needs then users may violate access control policies to get their jobs done \cite{base12}.  Previous work in developer-centered security \cite{pifz17} points to the presence of knowledge asymmetries.  As knowledge about the problem domain, threat model, and potential access control expertise is distributed, it is easy for different stakeholders to fail to appreciate the value of assets or the vulnerabilities a design affords should security requirements be miscommunicated or misinterpreted.

In this paper, we present a tool-supported technique for identifying knowledge asymmetries around access control based on asset and goal models.  By using simple Software and Requirements Engineering modelling languages that complement different design techniques, we provide boundary objects -- artifacts that are both flexible and robust enough to retain their identity when used by different stakeholders \cite{stgr89} -- that make access needs transparent to different stakeholders. This can contribute towards making stakeholder access control knowledge more symmetric.  We describe the related work upon which our technique is based in Section \ref{sect:relatedwork}, before presenting and illustrating its elements in Sections \ref{sect:approach} and \ref{sect:example} respectively, before concluding with directions for future work in Section \ref{sect:discussion}.

\section{Related Work}
\label{sect:relatedwork}

\subsection{Knowledge Asymmetries}

Previous work in developer-centered security \cite{pifz17} points to the presence of \emph{knowledge asymmetries}.  These are conditions created when different people or organisational units possess different stocks of knowledge \cite{roma08}.  Knowledge asymmetries can be an important source of innovation.  System design entails synthesizing different perspectives of a problem, so these asymmetries can be an opportunity to uncover tacit aspects of a problem \cite{fis00}.  

\begin{figure}[h!]
\centering
\includegraphics[scale=0.7]{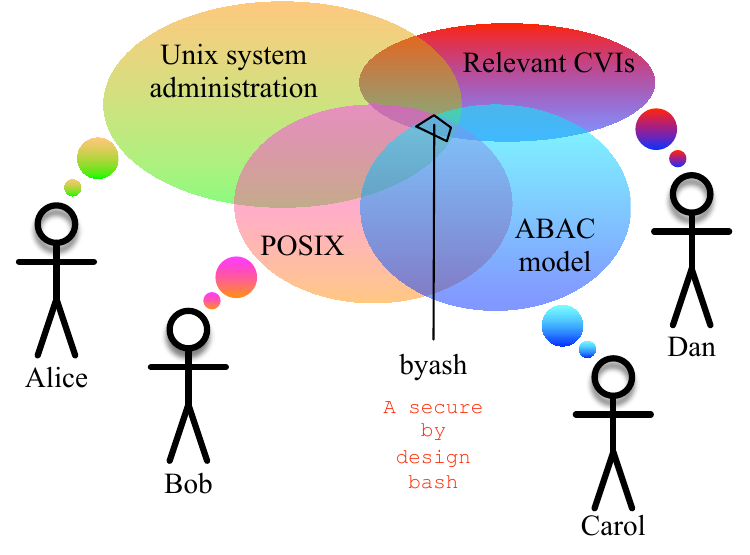}
\caption{Knowledge asymmetries in a ``secure by design'' shell}
\label{F:ka}
\end{figure}

Unfortunately, because knowledge about the problem domain, threat model, and potential access control expertise is distributed, it is easy for different stakeholders to fail to appreciate the value of assets or the vulnerabilities a design affords should security requirements be miscommunicated or misinterpreted.  For example, consider the design of a hypothetical ``secure by design'' \texttt{bash} shell:  \texttt{byash} (Bourne Yet Again Shell).  Figure \ref{F:ka} considers the different spaces of expertise that need to contribute to the  \texttt{byash} requirements.  Alice can provide the perspective of the Unix system administrator community that will use  \texttt{byash} on a day-to-day basis.  Because the shell itself will need to conform to the Portable Operating System Interface standard (POSIX) \cite{posix} to maintain compatibility and interoperability between operating systems, Bob - as a system programmer - will also need to provide some input.  It will also be desirable for  \texttt{byash} to implement data-centric security, so Carol's expertise on Attribute Based Access Control (ABAC) is required to ensure that policies are usable to the wide-range of end-users and developers.  Finally, Dan's experience carrying out Cyber Vulnerability Investigations (CVIs) on platforms that use the target operating system will provide insight into the  \texttt{byash} threat model necessary for security requirements capture.   

While this example sounds complex, it is comparatively trivial given the scale of some defence capabilities, and the myriad Defence Lines of Development expertise spaces that accompany them.  Consequently, boundary objects  -- artifacts that are both flexible and robust enough to retain their identity when used by different stakeholders \cite{stgr89} -- play an important role in creating spaces that intersect the spaces of expertise.  If these are sufficiently boundary spanning then they can decrease the negative effects of knowledge asymmetries \cite{roma08}.

\subsection{Model driven security}

Previous work in model driven security has examined the capture of access control intent using conventional modelling languages.  For example, SecureUML \cite{badl06} integrated access control into UML class diagrams for subsequent model transformation.  In this approach, roles (as subjects) have permissions for actions on resources, where roles could be groups or users, permissions are constraints, and actions could be atomic or composite. Authorisation constraints were modelled as Object Constraint Language (OCL) constraints on the ends of association. Building on this approach have been attempts to enforce access control policies based on such models, e.g. \cite{bdlj15}.  A limitation of this previous work has been the focus on model transformation.  Moreover, while such approaches are expressive, they assume the presence of a detailed design based on validated requirements.  Basin et al. \cite{bace11} also note that expressive modelling languages are problematic for managing consistency between different models.  They allude to the challenge of developing modelling languages expressive enough to capture policies, support formal analysis, and provide a basis for generating infrastructure to enforce or at least monitor policies.

\subsection{Security Requirements Engineering}

Previous work in Goal Oriented Requirements Engineering has considered requirements modelling for access control.  Giorgini et al. \cite{gmmz05} describe the modelling of access control permissions as predicates where an actor \emph{a} owns service \emph{s}, or actor \emph{a} has permission to use service \emph{s}. These are subsequently used by axioms to determine how permissions might cascade delegation chains or subgoals. This support was subsequently added to the Secure Tropos methodology \cite{gmz07}.  Similarly, Paja's STS-ml modelling language \cite{paj14} incorporates an authorisation view, which models the transfer and propagation of permissions between actors and allowed or prohibited read, modify, produce, or transfer operations.  STS-ml also incorporates complementary tool-support (STS-tool), which can verify the fulfilment of security requirements.  However, these approaches rely on extensions to social goal modelling that are already visually complex, which can hinder take-up by stakeholders \cite{mohe09}.

IRIS (Integrating Requirements and Information Security) is a process framework for designing usable and secure software \cite{fail18}. It is also supported by the CAIRIS (Computer Aided Integration of Requirements and Information Security) platform: an open-source platform for specifying, modelling, and validating requirements, security, and usability models.  CAIRIS can automatically validate and visualise several views of a system being specified.  These include asset models, class diagrams of assets based on the AEGIS method \cite{flech03}, and system goal models based on the KAOS goal modelling language \cite{lams09}.  IRIS and CAIRIS, together with asset and goal models, have been used to analyse and improve a security policy for a critical infrastructure company.  However, this approach did not account for the access control needs of stakeholders, and how this aligned with the security policy.

\section{Approach}
\label{sect:approach}

The technique entails simultaneously modelling access control needs and requirements using AEGIS asset models and KAOS goal models respectively.  This is complemented by an algorithm to spot potential issues as both models evolve.  Both modelling languages were chosen due to their expressiveness and, given the problems raised by \cite{mohe09}, low visual complexity.  This facilitates their use as boundary objects by stakeholders regardless of their expertise.

\subsection{Modelling access control needs with class diagrams}
\label{ss:model}

The approach for modelling access control needs is based on UML class diagrams, where a class represents an asset: someone or something of value, and its properties are one or more qualitative security properties, such as Confidentiality and Integrity; each property value is rated as $None$, $Low$, $Medium$, or $High$, where $None < Low < Medium < High$.  These properties are meant to stimulate dialogue between stakeholders about what constitutes $None$, $Low$, $Medium$, and $High$ values.  For example, disclosure of a low Confidentiality information asset might hinder day-to-day operation of the system, whereas tampering with a medium Integrity system asset may cause notable damage to operations.  It is not the role of the asset to specify what this hindrance or damage might be; this is the subject of complementary threat modelling or risk analysis.  An asset is also typed as a $System$, $Information$, or $People$.

\begin{figure}[h!]
\centering
\includegraphics[scale=0.5]{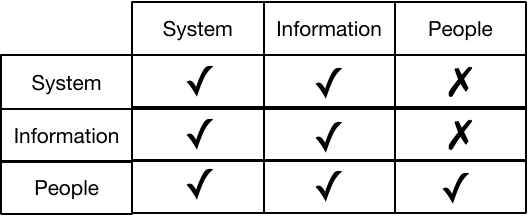}
\caption{Access rules between subjects (rows) and resources (columns)}
\label{fig:ar}
\end{figure}

In the technique, both the \emph{subject} requiring access, and the \emph{resource} accessed are assets.  However, as indicated in Figure \ref{fig:ar}, we restrict the types of assets that can request access to other assets.  For example, a person might read some information, but not vice-versa.  It may appear odd that information should be permitted to access resources but, during early stages of design, stakeholders might model some system that stores information as an information asset, or an information asset needs to access some resource because some out-of-scope system or person is handling that information, but the information-information access has some value.

\begin{figure}[h!]
\centering
\includegraphics[scale=0.5]{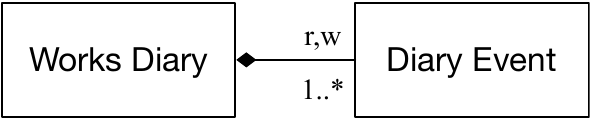}
\caption{A class diagram with access needs}
\label{fig:cd}
\end{figure}

Unlike existing approaches like \cite{badl06} which rely on OCL to model \emph{constraints} on access between classes, we take a simpler approach that entails modelling access \emph{needs}; such needs are easier to capture than constraints early in design.  We rely on the optional adornment of class association ends with information about the role played by the asset on the end of a relationship.  We exploit this feature by adorning ends with access control information.  For example, Figure \ref{fig:cd} models the relationship between a Works Diary and Diary Event; both are information assets and the former contains one or more of the latter.  The \emph{r,w} adornment on the tail end of the Works Diary-Diary Event association should be read as ``a works diary needs to read and write a diary event''.  Possible adornments for access needs are one or more of read (\emph{r}), write (\emph{w}), and interact (\emph{x}).  The choice of \emph{x} is based on the execute permission used in unix, but the term \emph{interact} allows for a range of affordances beyond execution.  Where \emph{r}, \emph{w}, or \emph{x} adornments are omitted from the end of an association, no access is assumed necessary, e.g. in Figure \ref{fig:cd}, diary events do not read, write, or interact with works diaries.  This approach is based on AEGIS, but where AEGIS considers asset value and requirements elicitation as synonymous, asset values are necessary but not sufficient for security requirements elicitation.  Asset modelling typically takes place too early in the design phase to elicit useful security requirements.  For example, setting low Confidentiality and Integrity properties for a Word Processor asset might indicate that the word processor is not used for sensitive work.  The values, however, tell us nothing about what the system needs to do to ensure these values are protected.

\subsection{Modelling access control policies with requirements}

Many organisations have multiple access control policies for disparate systems.  Although these may be maintained by security teams in isolation, these policies impinge on system goals operationalised by tasks individuals carry out.  To capture this impact, our technique directly aligns access control policies with access control requirements modelled in KAOS.  An access control policy captures the set of authorised and unauthorised interactions between assets.  Each access control policy statement is defined as $Subject$ $\times$ $Access$ $\times$ $Resource$ $\times$ $Permission$ where $Subject$ and $Resource$ are assets, $Access$ = $\{read, write, interact\}$ and $Permission$ = $\{allow,deny\}$, and each policy statement is associated with a single requirement.  The requirements should be precise enough to specify the conditions or capabilities the system needs to satisfy for the policy statements to hold.  Where this is not possible, the requirement would need to be subject to refinement.  Consequently, a complete access control policy should correspond with a complete specification describing the intent the system needs to satisfy for compliance with the policy.  
 
When user needs are met, the access needs in class diagrams should correspond with policy statements with an $allow$ permission.  However, the absence of access needs need not correspond with a $deny$ permission for the related subject and resource assets.  Other means exist for capturing the rationale for non-inclusion, and requirements for denying access may not be within the scope of analysis.  The approach allows the addition of requirements and policy statements if they are.

\subsection{Validating access control needs to identify asymmetries}

Access control can be validated with stakeholders by reviewing whether the desired access control needs are permitted by the access control policy.  Where these needs are absent between particular subjects and resources, designers should agree on whether the needs exist and, if not, requirements denying access should be specified.  The goal model aids transparency by helping stakeholders appreciate what an \emph{allow} or \emph{deny} means for some desired interaction.

\begin{algorithm}
  \SetKwInOut{Input}{Input}
  \small
  \SetAlgoLined
  \Input{$cmGraph$ - asset model graph}
  \KwData{$expandedNeeds$ - set of access need triples ($Subject$ $\times$ $Access$ $\times$ $Resource$) drawn from asset model , $aPolicyStmt$ - permitted policy statement, $dPolicyStmt$ - denied policy statement, $subjValue$ - subject security property value, $resValue$ - resource security property value}
  
  \SetKwProg{Fn}{Function}{ is}{end}
  \Fn{validationCheck($cmGraph$)}{
  
  $expandedNeeds$ $\leftarrow$ $\emptyset$;
  
  \While{$subj$ $needs$ $res$ $\leftarrow$ $cmGraph$} {
    \While{$acNeed$ $\leftarrow$ expandNeeds $needs$}{
      $expandedNeeds$ $\cup$ \{$subjt$,$acNeed$,$res$\};
    }
  } 
  
  \While{$subj$ $acNeed$ $res$ $\leftarrow$ $expandedNeeds$ }{
    $aPolicyStmt$ $\leftarrow$ allowedPolicyStatement $subj$ $acNeed$ $res$; 
    
    \eIf{$aPolicyStmt$ =  $\emptyset$}{
      $dPolicyStmt$ $\leftarrow$ deniedPolicyStatement $subj$ $acNeed$ $res$;
      
      \eIf{$dPolicyStmt$ =  $\emptyset$}{
        log\ 'Unauthorised access'\ $subj$\ $acNeed$\ $res$;
      }
      {
        log\ 'Undefined access'\ $subj$\ $acNeed$\ $res$;
      }
    }
    {
      $subjValue$ $\leftarrow$ confidentialityProperty $subj$;
      
      $resValue$ $\leftarrow$ confidentialityProperty $res$;

      \tcp{Check Simple Security Property}      
      \If{($resValue$ $>$ $subjValue$) $\land$ ($acNeed$ = read) }{
         log\ 'Potential no read-up violation'\ $subj$\ $acNeed$\ $res$;      
      }

      \tcp{Check *-Security Property}
      \If{($subjValue$ $>$ $resValue$) $\land$ ($acNeed$ = write) }{
         log\ 'Potential no write-down violation'\ $subj$\ $acNeed$\ $res$;     
      }
      
      $subjValue$ $\leftarrow$ integrityProperty $subj$;
      
      $resValue$ $\leftarrow$ integrityProperty $res$;   

      \tcp{Check Simple Integrity Property}
      \If{($resValue$ $>$ $subjValue$) $\land$ ($acNeed$ = write) }{
         log\ 'Potential no write-up violation'\ $subj$\ $acNeed$\ $res$;      
      }   
      
      \tcp{Check Integrity *-Property}
      \If{($subjValue$ $>$ $resValue$) $\land$ ($acNeed$ = read) }{
         log\ 'Potential no read-down violation'\ $subj$\ $acNeed$\ $res$;      
      }  
              
    }
  
  }
  \textbf{return};
}
\caption{Access control validation check}
\label{alg:acvc}
\end{algorithm}

Algorithm \ref{alg:acvc} was implemented in CAIRIS to warn of potential issues with access needs and the access control policy.  Where access needs are permitted, the policy statements are verified against security rules for particular security models.  For illustrative purposes, the rules used in this algorithm are drawn from the Bell-LaPadula \cite{bela73} and Biba \cite{biba75} security models, which consider the preservation of information Confidentiality and Integrity respectively.  Our approach makes no assumptions about what access control model or implementation will be enforced in the system being designed.  

The algorithm begins by breaking each association between assets into a triple of subject, roles, and resources where an access need is present (Line 3).  For example, the diagram in Figure \ref{fig:cd} yields the triple ($Works Diary$, ``$r,w$'', $Diary Event$); no triple would be present for the association between Diary Event and Works Diary because no access needs are present.  The needs are subsequently expanded (Line 4-6), such triples with multiple needs are transformed into multiple triples with single needs, e.g. ($Works  Diary$,``$r,w$'', $Diary Event$) expands to ($Works Diary$,``$r$'', $Diary Event$) and ($Works Diary$,``$w$'', $Diary Event$).

Each expanded access need is subsequently enumerated (Lines 8-38) to determine if the access control policy permits the need (Line 9).  If the need has not been explicitly permitted but the policy forbids this access then an unauthorised access warning is logged, otherwise an absent access warning is instead logged (Lines 9-16).  The logging of undefined access is intended to make this potential ambiguity transparent to designers and stakeholders, who can then review the access control policy, the access need, or both.  If the access need corresponds with a policy statement then we obtain the Confidentiality and Integrity properties associated with the subject and resources (Lines 18-19, 26-27) and compare these values based on access need and classes of violation in certain access control security models.  For example, where the Confidentiality property of a resource that needs to be read is greater than the subject then there is potentially a violation of the Bell-LaPadula Simple Security Property, i.e. no read-up \cite{bela73}.  Similarly, where the Integrity property of a subject is higher than the level of the resource it needs to read from then there is potentially a violation of the Biba Integrity *-Property, i.e. no read-down \cite{biba75}.

\section{Case Study: Identifying Access Control knowledge asymmetries in a PYRAMID component}
\label{sect:example}

\begin{figure*}[h!]
\centering
\includegraphics[scale=0.35]{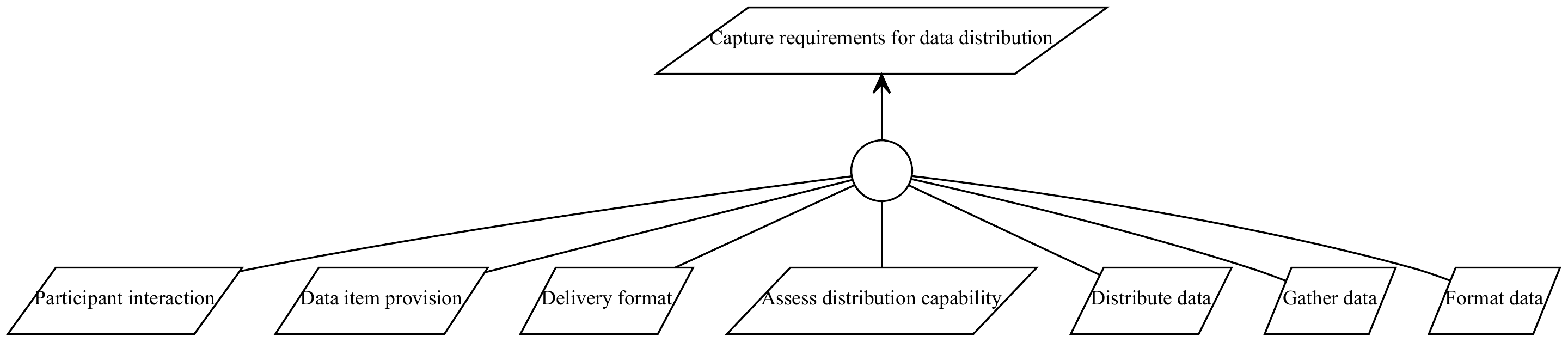}
\caption{KAOS goal model representing system requirements for data distribution}
\label{F:gm}
\end{figure*}

\subsection{PYRAMID overview}

We illustrate this technique by finding potential access control knowledge asymmetry when a pre-existing avionics software component is integrated into a hypothetical military air system\footnote{CAIRIS model and reviewer instructions for interacting with it available at \url{https://github.com/failys/icmcis2025_paper_model}}.  The component's specification is based on the Pyramid Reference Architecture (PRA).  PRA is an open, air system, reference architecture that is exploiting and execution platform independent \cite{pyramid}.  By providing an open architecture approach and encouraging systematic software reuse, PYRAMID aims to make avionic system design and procurement more affordable, capable, and adaptable.

To achieve PYRAMID's goals, the PRA alone is insufficient for developing PYRAMID components.  Exploiting programmes building air systems need to determine system and security requirements, while suppliers need to develop the software components based on their interpretation of the PRA.  When developing these components, suppliers may not have visibility of system and security requirements.  While requirements and security analysis may precede component development, this may not be the case where components are reused from a pre-existing air system.  As such, the ability to identify potential security issues before a component is integrated into an aircraft's broader software system could cast light on unwarranted assumptions that might otherwise be missed until much later.

The reused component is based on the \emph{Data Distribution} component specification within the PRA \cite{pyramidAA}.  The component's role is to prepare and distribute data between participants in a data exchange.  The component is agnostic to detail about the software and hardware platform it runs on, or the type of middleware used.  

\begin{figure*}[h!]
\centering
\includegraphics[scale=0.75]{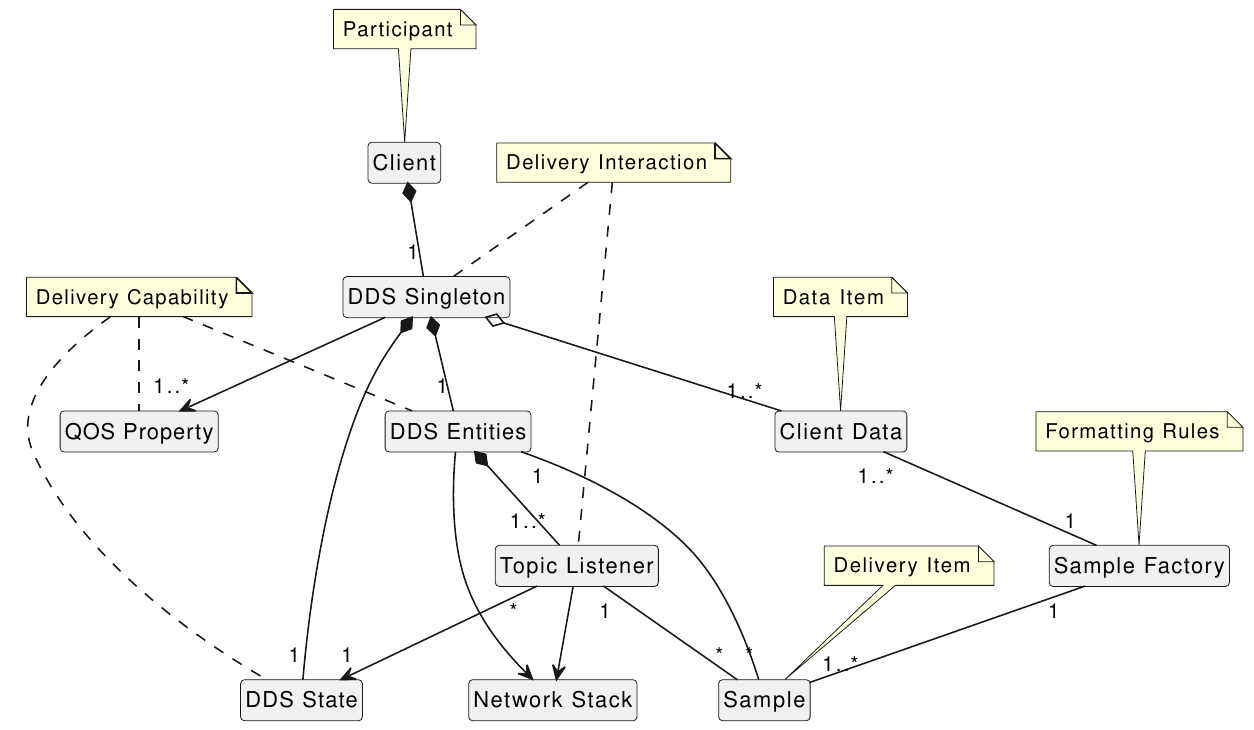}
\caption{Legacy Data Distribution class model}
\label{F:lcm}
\end{figure*}

A component implementing this specification was developed in C++ using Data Distribution Service (DDS)  middleware; DDS is a data-centric middleware protocol and Application Programming Interface (API) standard, based on a publisher-subscriber metaphor \cite{dds}.  This implementation is used in a pre-existing, legacy air system  The key elements of this component are illustrated in Figure \ref{F:lcm}.  Clients use an API for mission specific data distribution operations, e.g. reading and writing messages.  Behind the API is a singleton \cite{creational} object that encapsulates DDS entities such as publishers, subscribers, readers, and writers, and state specific information.  The DDS entities are configured at run-time using Quality of Service (QoS) properties set externally, i.e. through configuration files.  Client data passed into the API is translated into DDS sample data and vice-versa using a factory \cite{creational} object.  API operations that map to DDS publication operations are directly handled by the singleton object.  However, the receipt of DDS samples is handled by listener classes, which update the state specification information.  This state is available from ``read'' specific API operations.  

As Figure \ref{F:lcm}, there is alignment between the classes in this detailed design and the entities in the PRA.

\subsection{Modelling access control needs}

\begin{figure}[h!]
\centering
\includegraphics[scale=0.5]{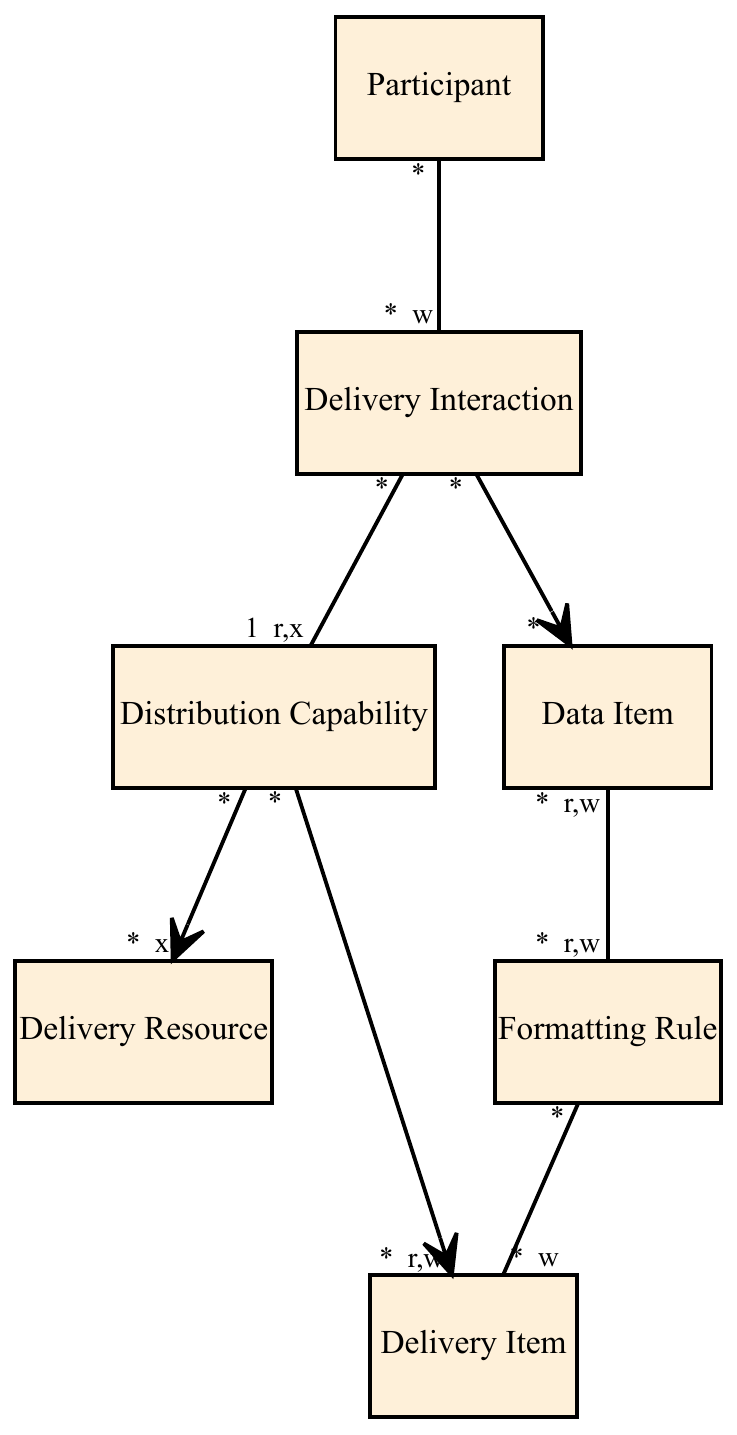}
\caption{Data Distribution component asset model}
\label{F:cam}
\end{figure}

Figure \ref{F:cam} shows the resulting asset model of these component elements.  The security properties for each asset are described in Table \ref{T:asp}.  The value for the properties is based on the potential impact of a breach at an associated classification tier.  For example, a Low value is associated with compromise, loss, or disclosure of the asset at an OFFICIAL classification \cite{gscp}, which is the indicative security classification for the component in the PRA.  

\begin{table}[]
\centering
\begin{tabular}{@{}lll@{}}
\toprule
\textbf{Asset}       & \textbf{Confidentiality} & \textbf{Integrity} \\ \midrule
Data Item            & Low          & Medium          \\
Distribution Capability  & None  & Medium          \\
Delivery Interaction & Low          & Medium          \\
Delivery Item        & Low          & Medium          \\
Delivery Resource    & None        & None          \\
Formatting Rule      & None          & Medium          \\
Participant          & None          & Low    \\ \midrule     
\end{tabular}
\caption{Data Distribution asset security properties}
\label{T:asp}
\end{table}

\subsection{Modelling system requirements and access control policy statements}
The security considerations within the component specification note that access to data should be restricted to ensure different classifications of communication remain separate and not inappropriately directed.  The PRA does not stipulate access control policies for this component, but these are implied within the component specification.  Consequently, based on the specification responsibilities, subject matter semantics, design rationale and considerations, and safety and security considerations, seven system requirements were elicited.  As indicated in Figure \ref{F:gm}, these requirements were a refinement of the  \emph{Capture requirements for data distribution} requirement within the component specification.  The access control policies for the component entities associated with these requirements are specified in Table \ref{T:ps}.

\begin{table*}[]
\centering
\begin{tabular}{@{}lllll@{}}
\toprule
\textbf{Requirement}                  & \textbf{Subject}     & \textbf{Access Type} & \textbf{Resource}    & \textbf{Permission} \\ \midrule
Participant interaction        & Participant          & write                & Delivery Interaction & allow               \\
Data item provision            & Participant          & read                 & Data Item            & allow               \\
Delivery format                & Formatting Rule      & read                 & Data Item            & allow               \\
Assess distribution capability & Delivery Interaction & read                 & Distribution Capability  & allow  \\
Distribute data                & Distribution Capability  & write                & Delivery Item        & allow               \\
Gather data                    & Delivery Interaction & read                 & Data Item            & allow               \\
Format data                    & Formatting Rule      & write                & Delivery Item        & allow               \\  \bottomrule
\end{tabular}
\caption{Data Distribution policy statements}
\label{T:ps}
\end{table*}

\subsection{Validating access needs}

\begin{table}[]
\centering
\begin{tabular}{@{}ll@{}}
\toprule
\textbf{Security rule} & \textbf{Violation}\\ \midrule
Simple Security Property & Y \\
*-Property  & N \\
Simple Integrity Property  & Y  \\
Integrity *-Property  & N \\ 
Absent policies & Y \\ \midrule
\end{tabular}
\caption{Security rule validation check results}
\label{T:srv}
\end{table}

Running the model validation algorithm yielded 8 warnings; the types of warning are summarised in Table \ref{T:srv}.  

Six of the warnings were due to absent policy statements where access was expected in the component implementation, but not specified in the access control policy associated with the system requirements.  For example, \emph{Distribution Capability} objects need to read \emph{Delivery Items} within the component implementation, i.e. to handle topic samples the participant might have subscribed to.  However, the policy statement for the component indicates that \emph{Delivery Capability} objects can only write \emph{Delivery Items}.  This warning identifies potential asymmetries of knowledge between the programme and supplier around which entities should be responsible for DDS publication and subscription; these need to be resolved before updating the policy statements.

The violation of the Simple Security Property occurred because \emph{Formatting Rule} objects need read access to \emph{Data Items}, but the Confidentiality property of the resource is higher than the subject.  While the PRA implies that \emph{Formatting Rule} is a static collection of rules, the component implements it using a collection of factory methods \cite{creational}, where the appropriate \emph{Delivery Item} object is created based on the type of \emph{Data Item} parameter.  While the policy stipulates that \emph{Formatting Rule} should be able to write \emph{Delivery Item} and read \emph{Data Item} objects, the component implementation needs to interact with the \emph{Formatting Rule} factories.  Because this is not stipulated as a policy statement, the warning should promote discussion between the programme team and supplier around expectations for formatting data before and after data distribution.

The violation of the Simple Integrity Property occurred because \emph{Participants} need write access to \emph{Delivery Interactions}, but the Integrity value of the resource is higher than that of the subject.  From the perspective of the supplier, the \emph{Participant} Integrity value is no higher than that implied within the PRA, and its corruption is not considered within scope.  However, for the exploiting programme, failing to protect \emph{Delivery Interactions} from tampering could lead to interactions at a lower-level of classification tainting interactions at a higher-level.  This situation could have arisen because the component had originally been deployed in a legacy environment where, for several reasons, such protections were unnecessary.

\section{Discussion and Conclusion}
\label{sect:discussion}

In this paper, a tool-supported technique was presented for identifying knowledge asymmetries around access control based on asset and goal models.  The technique does not replace comprehensive approaches for capturing security requirements, but could be plugged into such approaches.  As such, the paper makes three contributions.

First, we show how simple, conventional models that may already exist shed light on access control issues that might otherwise remain hidden; this can make access control more transparent leading to fewer knowledge asymmetries related to access control.  While the worked example is simple enough that the problems found could be spotted by careful inspection alone, such issues become harder to spot as models grow.  Our work also complements other participative design or threat modelling techniques.  For example, we could use outputs from premortems \cite{falp121} in threat modelling workshops to elicit access needs, validate asset values, or review access control requirements.  

Second, this work provides a view of what ``shifting access control left'' looks like.  Previous work \cite{faia17} has already shown how processes and tools akin to those in this report support the provision of machine-readable design models that can be incorporated into Continuous Integration / Continuous Development (CI/CD) pipelines.  Moreover, although the tooling that supported this approach was implemented in CAIRIS, there is no reason it could not be implemented in other modelling or system requirements/architecture tools that facilitated interoperability with other tools, i.e. by supporting an API.  Doing so ensures that thinking about access control is productive and does not inhibit other design or innovation practices.

Finally, we presented a validation approach, which is potentially extensible and doesn't preclude the specification of more elaborate policies, based on different access control models.  For example, the initial loop in Algorithm 1 could be refined to incorporate access needs resulting from role or object hierarchies expressed using class inheritance; this would allow needs associated with a Role-Based Access Control (RBAC) policy to be captured.  Similarly, if class attributes were also captured then fine-grained needs could be evaluated against ABAC models, i.e. those associated with data-centric security policies.  Moreover, while the algorithm does not traverse the asset model, a similar approach could be employed to \cite{fsss20}, where the graph could be traversed to obtain a set of asset association traces that could subsequently analysed for more elaborate taint checking, or modelling the impact of delegating subgoals similar to \cite{gmmz05}.  

For future work, we will conduct a more comprehensive validation of this technique with different access control models, participative design, and threat modelling techniques for eliciting access control requirements, and improving knowledge asymmetries.  We will also further improve the validation algorithm, including making greater use of usability and security concepts supported by IRIS to identify context of use nuances with access control implications.  

\section*{Acknowledgements}
This document is an overview of UK MOD sponsored research. The contents of this document should not be interpreted as representing the views of the UK MOD, nor should it be assumed that they reflect any current or future UK MOD policy.

\textcopyright~ Crown copyright (2024), Dstl. This information is licensed under the Open Government Licence v3.0. To view this licence, visit \url{https://www.nationalarchives.gov.uk/doc/open-government-licence/version/3}. Where we have identified any third party copyright information you will need to obtain permission from the copyright holders concerned. Any enquiries regarding this publication should be sent to: Dstl.

\balance
\bibliographystyle{IEEEtran}
\bibliography{manuscript}

\begin{thebibliography}{10}
\providecommand{\url}[1]{#1}
\csname url@samestyle\endcsname
\providecommand{\newblock}{\relax}
\providecommand{\bibinfo}[2]{#2}
\providecommand{\BIBentrySTDinterwordspacing}{\spaceskip=0pt\relax}
\providecommand{\BIBentryALTinterwordstretchfactor}{4}
\providecommand{\BIBentryALTinterwordspacing}{\spaceskip=\fontdimen2\font plus
\BIBentryALTinterwordstretchfactor\fontdimen3\font minus
  \fontdimen4\font\relax}
\providecommand{\BIBforeignlanguage}[2]{{%
\expandafter\ifx\csname l@#1\endcsname\relax
\typeout{** WARNING: IEEEtran.bst: No hyphenation pattern has been}%
\typeout{** loaded for the language `#1'. Using the pattern for}%
\typeout{** the default language instead.}%
\else
\language=\csname l@#1\endcsname
\fi
#2}}
\providecommand{\BIBdecl}{\relax}
\BIBdecl

\bibitem{sbd}
{Ministry of Defence}, ``{Secure by Design: Secure from the start},''
  \url{https://www.digital.mod.uk/secure-by-design/secure-from-the-start},
  August 2024.

\bibitem{wron24}
K.~Wrona, ``{Towards Data-Centric Security for NATO Operations},'' in
  \emph{{Digital Transformation, Cyber Security and Resilience}}, T.~Tagarev
  and N.~Stoianov, Eds.\hskip 1em plus 0.5em minus 0.4em\relax Springer, 2024,
  pp. 75--92".

\bibitem{base12}
S.~Bartsch and M.~A. Sasse, ``Guiding decisions on authorization policies: A
  participatory approach to decision support,'' in \emph{Proceedings of the
  27th Annual ACM Symposium on Applied Computing}.\hskip 1em plus 0.5em minus
  0.4em\relax ACM, 2012, p. 1502–1507.

\bibitem{pifz17}
O.~Pieczul, S.~Foley, and M.~E. Zurko, ``Developer-centered security and the
  symmetry of ignorance,'' in \emph{Proceedings of the 2017 New Security
  Paradigms Workshop}.\hskip 1em plus 0.5em minus 0.4em\relax ACM, 2017, pp.
  46--56.

\bibitem{stgr89}
S.~L. Star and J.~R. Griesemer, ``Institutional ecology, "translations" and
  boundary objects: Amateurs and professionals in berkeley's museum of
  vertebrate zoology, 1907-39,'' \emph{Social Studies of Science}, vol.~19, pp.
  387--420, 1989.

\bibitem{roma08}
M.~Ronko and M.~Makela, ``{Asymmetries of Knowledge Between Engineering and
  Marketing in Software Product Development},'' in \emph{Proceedings of the
  2018 European Conference on Information Systems}.\hskip 1em plus 0.5em minus
  0.4em\relax Association for Information Systems, 2008.

\bibitem{fis00}
G.~Fischer, ``Symmetry of ignorance, social creativity, and meta-design,''
  \emph{Knowledge-Based Systems}, vol.~13, no.~7, pp. 527--537, 2000.

\bibitem{posix}
{The Open Group and IEEE}, ``{IEEE Std 1003.1 - 2024 Edition},''
  \url{https://pubs.opengroup.org/onlinepubs/9799919799}, August 2024.

\bibitem{badl06}
D.~Basin, J.~Doser, and T.~Lodderstedt, ``Model driven security: From uml
  models to access control infrastructures,'' \emph{ACM Transactions on
  Software Engineering and Methodology}, vol.~15, no.~1, pp. 39--91, Jan. 2006.

\bibitem{bdlj15}
J.~Bogaerts, M.~Decat, B.~Lagaisse, and W.~Joosen, ``Entity-based access
  control: Supporting more expressive access control policies,'' in
  \emph{Proceedings of the 31st Annual Computer Security Applications
  Conference}.\hskip 1em plus 0.5em minus 0.4em\relax ACM, 2015, p. 291–300.

\bibitem{bace11}
D.~Basin, M.~Clavel, and M.~Egea, ``A decade of model-driven security,'' in
  \emph{Proceedings of the 16th ACM Symposium on Access Control Models and
  Technologies}.\hskip 1em plus 0.5em minus 0.4em\relax ACM, 2011, pp. 1--10.

\bibitem{gmmz05}
P.~{Giorgini}, F.~{Massacci}, J.~{Mylopoulos}, and N.~{Zannone}, ``Modeling
  security requirements through ownership, permission and delegation,'' in
  \emph{Proceeedings of the 13th IEEE International Conference on Requirements
  Engineering}.\hskip 1em plus 0.5em minus 0.4em\relax IEEE, 2005, pp.
  167--176.

\bibitem{gmz07}
P.~Giorgini, H.~Mouratidis, and N.~Zannone, ``{Modelling Security and Trust
  with Secure Tropos},'' in \emph{Integrating Security and Software
  Engineering}.\hskip 1em plus 0.5em minus 0.4em\relax Idea Group, 2007.

\bibitem{paj14}
E.~Paja, ``{STS: A Security Requirements Engineering Methodology for
  Socio-Technical Systems},'' Ph.D. dissertation, Universit\`{a} degli Studi di
  Trento, 2014.

\bibitem{mohe09}
D.~L. Moody, P.~Heymans, and R.~Matulevicius, ``Improving the effectiveness of
  visual representations in requirements engineering: An evaluation of i*
  visual syntax,'' in \emph{{Proceedings of the 17th IEEE International
  Requirements Engineering Conference}}.\hskip 1em plus 0.5em minus 0.4em\relax
  IEEE, 2009, pp. 171--180.

\bibitem{fail18}
S.~Faily, \emph{{Designing Usable and Secure Software with IRIS and
  CAIRIS}}.\hskip 1em plus 0.5em minus 0.4em\relax Springer, 2018.

\bibitem{flech03}
I.~Fl\'{e}chais, M.~A. Sasse, and S.~M.~V. Hailes, ``Bringing security home: a
  process for developing secure and usable systems,'' in \emph{Proceedings of
  the 2003 New Security Paradigms Workshop}.\hskip 1em plus 0.5em minus
  0.4em\relax ACM, 2003, pp. 49--57.

\bibitem{lams09}
A.~van Lamsweerde, \emph{{Requirements Engineering: from system goals to UML
  models to software specifications}}.\hskip 1em plus 0.5em minus 0.4em\relax
  John Wiley \& Sons, 2009.

\bibitem{bela73}
D.~E. Bell and L.~J. LaPadula, ``{Secure Computer Systems: Mathematical
  Foundations},'' {MITRE}, 2547, 1974.

\bibitem{biba75}
K.~J. Biba, ``{Integrity Considerations for Secure Computer Systems},''
  {MITRE}, Tech. Rep. 3153, 1975.

\bibitem{pyramid}
{Ministry of Defence}, ``{PYRAMID Exploiter's Pack Version 4.1},''
  \url{https://assets.publishing.service.gov.uk/media/651167922f404b0014c3d850/PYRAMID_Exploiter_s_Pack_Main_Document_Issues_4.1.pdf},
  September 2023.

\bibitem{pyramidAA}
------, ``{PYRAMID Exploiter's Pack Version 4.1: Annex A - PRA Description
  Document Issue 4.1},''
  \url{https://assets.publishing.service.gov.uk/media/651167922f404b0014c3d850/PYRAMID_Exploiter_s_Pack_Main_Document_Issues_4.1.pdf},
  September 2023.

\bibitem{dds}
{Object Management Group}, ``{DDS Foundation Portal},''
  \url{https://www.dds-foundation.org}, January 2023.

\bibitem{creational}
E.~Gamma, R.~Helm, R.~Johnson, and J.~Vlissides, ``{Creational Patterns},'' in
  \emph{{Design Patterns: Elements of Reusable Object-Oriented
  Software}}.\hskip 1em plus 0.5em minus 0.4em\relax Addison-Wesley, 1994.

\bibitem{gscp}
{Cabinet Office}, ``{Government Security Classifications Policy},''
  \url{https://www.gov.uk/government/publications/government-security-classifications/government-security-classifications-policy-html},
  August 2024.

\bibitem{falp121}
S.~Faily, J.~Lyle, and S.~Parkin, ``Secure system? challenge accepted: Finding
  and resolving security failures using security premortems,'' in
  \emph{Designing Interactive Secure Systems: Workshop at British HCI 2012},
  2012.

\bibitem{faia17}
S.~Faily and C.~Iacob, ``Design as code: Facilitating collaboration between
  usability and security engineers using cairis,'' in \emph{2017 IEEE 25th
  International Requirements Engineering Conference Workshops (REW)}, 2017, pp.
  76--82.

\bibitem{fsss20}
S.~Faily, R.~Scandariato, A.~Shostack, L.~Sion, and D.~Ki-Aries,
  ``Contextualisation of data flow diagrams for security analysis,'' in
  \emph{Graphical Models for Security}.\hskip 1em plus 0.5em minus 0.4em\relax
  Springer, 2020, pp. 186--197.

\end{thebibliography}
\end{document}